\documentclass[10pt]{llncs}
\usepackage{makeidx}  
\usepackage{cite}
\pdfoutput=1

\usepackage{color}
\usepackage{epic,eepic,amsfonts,amsmath,latexsym,amssymb,url}
\usepackage{ifthen,graphics,epsfig}
\usepackage[english]{babel}
\usepackage{times}
\usepackage{setspace}
\usepackage{multirow}
\usepackage{empheq}
\usepackage{graphicx}
\usepackage{framed}
\usepackage{fancyhdr}
\usepackage{float}
 \usepackage{caption}
 \usepackage{subcaption}
\usepackage{ tipa }

 \definecolor{OliveGreen}{RGB}{85,107,47}

\begin{document}
\newlength {\squarewidth}
\renewenvironment {square}
{
\setlength {\squarewidth} {\linewidth}
\addtolength {\squarewidth} {-12pt}
\renewcommand{\baselinestretch}{0.75} \footnotesize
\begin {center}
\begin {tabular} {|c|} \hline
\begin {minipage} {\squarewidth}

\medskip
}{
\end {minipage}
\\ \hline
\end{tabular}
\end{center}
}



\newcommand{\toto}{xxx}

\newcommand{\black}{{\em black}}
\newcommand{\red}{{\em red}}
\newcommand{\yellow}{{\em yellow}}
\newcommand{\brown}{{\em brown}}
\newcommand{\blue}{{\em blue}}
\newcommand{\orange}{{\em orange}}
\newcommand{\moving}{{\em moving}}
 
\newenvironment{proofT}{\noindent{\bf
Proof. }} {\hspace*{\fill}$\Box_{Theorem~\ref{\toto}}$\par\vspace{1mm}}
\newenvironment{proofL}{\noindent{\bf
Proof. }} {\hspace*{\fill}$\Box_{Lemma~\ref{\toto}}$\par\vspace{1mm}}
\newenvironment{proofC}{\noindent{\bf
Proof. }} {\hspace*{\fill}$\Box_{Corollary~\ref{\toto}}$\par\vspace{1mm}}
\newif\ifshowcomments
\showcommentstrue

\ifshowcomments
\newcommand{\mynote}[2]{\fbox{\bfseries\sffamily\scriptsize{#1}}
 {\small$\blacktriangleright$\textsf{\emph{#2}}$\blacktriangleleft$}}
\else
\newcommand{\mynote}[2]{}
\fi
\newcommand{\todo}[1]{\textcolor{red}{\mynote{GADL-TODO}{#1}}}
 \newcommand{\nota}[1]{\textcolor{blue}{\mynote{GADL-Nota}{#1}}}

\newcounter{linecounter}
\newcommand{\linenumbering}{\ifthenelse{\value{linecounter}<10}
{(0\arabic{linecounter})}{(\arabic{linecounter})}}
\renewcommand{\line}[1]{\refstepcounter{linecounter}\label{#1}\linenumbering}
\newcommand{\resetline}[1]{\setcounter{linecounter}{0}#1}
\renewcommand{\thelinecounter}{\ifnum \value{linecounter} >
9\else 0\fi \arabic{linecounter}}

\newcommand{\vir}[1]{``#1''}
\newcommand{\unifleftarrow}{$\xleftarrow{u}$}
 
 \newcommand{\blank}{{\em \textcrb}}
 
\newcounter{mycount}
\newcommand\myprob[2]{%
  \stepcounter{mycount}
  \par\noindent Problem\ \themycount  -
   {\bf #1:}
   #2\par
}

\title{\bf   Robots with Lights: Overcoming Obstructed Visibility Without Colliding }


\author{G. A.  Di Luna \inst{1}
\and P. Flocchini \inst{2}
\and S. Gan Chaudhuri  \inst{3}
\and N. Santoro\inst{4}
\and G. Viglietta \inst{2}
}

\institute{ 
\scriptsize
Dipartimento di Ingegneria Informatica, Automatica e Gestionale Antonio Ruberti, Universit\`a degli Studi di Roma ``La Sapienza'', Rome, Italy, {\tt diluna@dis.uniroma1.it} \and 
School of Electrical Engineering and Computer Science, University of Ottawa, Ottawa ON, Canada, {\tt flocchin@site.uottawa.ca, viglietta@gmail.com} \and
Department of Information Technology, Jadavpur University, Kolkata, India, {\tt srutiganc@it.jusl.ac.in}
\and  School of Computer Science, Carleton University, Ottawa ON, Canada, {\tt santoro@scs.carleton.ca}
}

\thispagestyle{empty}
\maketitle
\begin{abstract}

{\em Robots with lights} is a model of autonomous mobile
computational entties
  operating in the plane
   in Look-Compute-Move cycles:
each agent has   an externally  visible   light
which can  assume colors from a fixed set; the
  lights  are persistent (i.e., the color is not erased at the end
of a cycle), but otherwise
the agents are {oblivious}.
The investigation of computability in this model, initially suggested  
by Peleg,
is under way, and several results have been recently established.
In these investigations, however,
an agent is assumed to be capable
to see through another agent. \\
In this paper we start the study of computing
when
visibility is obstructable, and investigate
the most basic problem for this setting, {\em Complete Visibility}:
  The agents must reach
   within finite time a configuration where they can all see each  
other and terminate. 
We do not make any assumption  on   a-priori knowledge  of  the number  
of agents, on rigidity of movements nor on chirality.
   The local  coordinate system of an agent may change at each  
activation.    
Also, by definition of lights,  an agent can  communicate and   
remember only a constant number of bits in each cycle. 
In spite of these weak conditions, we  prove that  {\sc Complete  
Visibility}   is always solvable, even   in the {\em asynchronous}  
setting,
without collisions   and using  a small constant number of colors.
The proof is constructive.  
We  also show how to extend our protocol  for  {\sc Complete  
Visibility} so that,
with the same  number of
colors, the agents  solve   the (non-uniform) {\sc Circle Formation}  
problem with obstructed visibility.
\end{abstract}

\section{Introduction}

   \subsection{Framework}

   In the  traditional model of distributed computing by mobile
entities  in the plane, called {\em robots} or {\em agents}, each entity
   is modelled as a point;  it is provided with a local coordinate
system (not necessarily consistent with that of the other agents); it
has
   sensorial capabilities, called {\em vision}, enabling it to determine
   the position (within its own coordinate system) of   the other agents.
   The agents are anonymous, they are indistinguishable,  and they
execute the same code.

Agents operates in   {\em Look-Compute-Move} cycles: when becoming active,
    an agent uses its sensing capabilities to get a snapshot of its
surroundings (Look),  then this snapshot is used to compute 
a destination point (Compute), and finally it
moves towards this   destination  (Move); after that, the
agent becomes inactive.
    In the majority of investigations, the agents are assumed to be
oblivious: at the beginning of each cycle, an agent has no
recollection of
   its  past observations and computations \cite{FlPS12}.
Depending on the assumptions on   the activation schedule  and the
duration of the cycles, three main
settings are identified. In the {\em fully-synchronous} setting, all
agents are activated simultaneously, and
each cycle is instantaneous. The {\em semi-synchronous} setting is
like the fully synchronous one
except that he set of agents to be activated is chosen by an
adversary, subject only to a fairness restriction: each agent will
be activated infinitely often. In the {\em asynchronous} setting,
there is no common notion of time,  and no assumption is made on
timing of activation,
other than fairness,
nor on the duration of each computation and movement, other than it is finite.

   Vision and mobility provide the agents with {\em stigmergy}, enabling
    the agents to  communicate and coordinate their actions   by moving
and sensing their relative positions.
    The agents  are otherwise assumed to be {\em silent}, without any
   means of explicit direct  communication \cite{FlPS12}.
   This restriction enables   deployment in extremely harsh
environments where communication is not possible, i.e an underwater
deployment or a military scenario where wireless communication are
impossible or can be jammed. Nevertheless, in many  other situations
it is possible to assume the availability of some sort of direct
communication. The theoretical interest is obviously for weak
communication capabilities.

A model employing a  weak explicit communication mechanism is that of
{\em robots with lights}:
   in this model,
    each agent is provided with  a local externally  visible {\em  light},
which can can assume colors from a fixed set;  the
agents explicitly communicate  with each other using these lights
\cite{DasFPSY12,DasFPSY14,EfP07,FlSVY13,Peleg2005,Vi13}.
In this model,
the lights  are persistent (i.e., the color is not erased at the end
of a cycle), but otherwise
the agents are {oblivious}.

The classical model of silent entities and  the more recent model of
entities with visible lights share a common assumption, that  {\em
visibility is unobstructed}. That is,  three or more collinear agents
   are assumed
to be mutually visible.
It can be easily argued against such an assumption, and for the
importance of investigating computability  when  visibility is
obstructed by
presence of the agents: given three collinear agents, the one in the
middle  blocks the visibility  between  the other two and they cannot see
each other.

Nothing is known   on computing with {\em obstructed visibility}
except for the investigations on the so-called \emph{fat agents}
model, where agents are not points but unit discs, and collisions are
   allowed\footnote{In pointilinear models, collisions  create
unbreakable symmetries; thus, unless  this is the required outcome of
the problem,
their avoidance is required by all solution protocols.
In addition, in  real world implementations,   collisions (e.g.,  of two
quad copters)
  may have unpredictable outcomes that might be better  avoided.}
and can be  used as an explicit computational tool.
(e.g.,
\cite{BoKF12,CzGP09,AgGM13}); and
for
   the study of uniformly  spreading agents operating in a one dimensional space
(i.e., on a line) \cite{CoP08}.
In this paper we start to fill this void, and focus on agents with
visible lights in presence of    obstructed visibility.

The problem we investigate is perhaps the most basic in a situation of
obstructed visibility, and it is the one of
the agents  reaching a configuration of complete un-obstructeded
visibility. More precisely, this problem, that we shall call {\sc
Complete Visibility},
   requires the agents, starting from an arbitrary initial
configuration where  they are in distinct points but might be unable
to see
everybody and  might not know the total number of agents\footnote{The
actual number of agents may be unknown for several reasons; e.g.,
  if the deployment of agents has been done by an airplane, a
subset of agents may be lost or destroyed during the landing
process.}, to reach within finite time a configuration in which every
agent is in a distinct location from which it can see all other
agents, and no longer move.

Among the configurations that achieve complete visibility, a special
class is that where all agents are on the perimeter of a  circle (not
necessarily equally spaced). The problem of forming any such
a configuration is called {\sc Circle Formation}
and it has been extensively studied
both in the classical model
of silent agents and in the ones with visible lights
   (e.g.,  
\cite{DiLaPe08,SuS90,Defago2008,DatDGM13,Katreniak2005}).
Unfortunately, none of these
investigations consider obstructed visibility, and their algorithms do
not work in the setting considered here.

\subsection{Our Contributions}

In this paper we study solving  {\sc Complete Visibility} by robots  
with lights. That is, we consider autonomous and anonymous agents,     
each endowed with a visible light that can assume a constant number of  
persistent colors, that are otherwise oblivious, and whose visibility  
is obstructed by other agents in the line of sight; and we investigate  
under what conditions they can solve {\sc Complete Visibility} and at  
what cost (i.e., how many colors).

We do not make any assumptions  on  a-priori knowledge  on  the number  
of agents,  nor on agreement on  coordinate systems, unit of distance  
and  chirality; actually,  the local  coordinate system of an agent  
may change at each activation.   Neither we make any assumption on  
rigidity of movements;
that is, a move  may
be stopped by an adversary  before the
agent reaches its destination; the only constraint is that, if interrupted
before reaching its destination, the agent moves  at least a
  minimum distance  $\delta>0$
(otherwise, no destination can ever be reached).  Also, by definition  
of lights,  an agent can  communicate and  remember only a constant  
number of bits in each cycle.

In spite of these weak conditions, we  prove that  {\sc Complete  
Visibility}   is always solvable, even   in the {\em asynchronous}  
setting,
without collisions   and using  a small constant number of colors.
The proof is constructive.
We first design  a  protocol that achieves complete visibility     
with six  colors under a semi-synchronous scheduler.
We then show how to   transform it into an asynchronous algorithm  
with only  four  additional colors.
We also show how to extend the protocol so that, under the same weak  
conditions and without increasing the number of
colors, the agents can position themselves  on the perimeter of a  circle.
   In other words, we  also show how to solve the (non-uniform) {\sc  
Circle Formation} problem with obstructed visibility.

Due to lack of space, some of the proofs are sketched and some omitted.

\section{Model and Definitions\label{model}}

Consider  a set of mobile anonymous agents $\mathcal{A}:\{a_1,a_2,..,a_n\}$. 
Each agent $a_i$ has a persistent state variable $s_{i}$, which may assume any value 
in a finite set of {\em colors}  $C$. 
We denote by  $x_i(t) \in \mathbb{R}^{2}$ the position occupied by  agent 
$a_i$ at time $t$  expressed in some global coordinate system (used only for description purposes, and unknown to the agents);  when no ambiguity arises, we omit the indication of time. 
A {\em configuration} $\mathcal{C}$ is a set of $n$ tuples in $C \times \mathbb{R}^{2}$ each defining the position and color of an agent; let   $\mathcal{C}_{t}$ denote the configuration at time $t$.

Each agent $a_i$ has its own system of coordinates centered in itself, which does not necessarily    agree 
with those of the other agents, i.e. there is no common unit of measure and 
not common notion of clockwise orientation.
Agents $a_i$ and $a_j$ are visible to each other at time $t$
 if and only if the segment  $\overline{x_i(t)  x_j(t)}$ does not contain any other agents.  
Let  ${\mathcal{C}_t[a_{i}]} $ denote the set of the positions   and   colors of the agents visible 
to $a_i$     time $t$. We shall call such a set {\em local view}.  A configuration
 $\mathcal{C}$ is said to be {\em obstruction-free} 
 if $\forall a_i \in \mathcal{A}$ we have $|\mathcal{C}[a_i]|=n$; that is, if all agents can see each other.  
 Two agents $a_i$ and $a_j$ are said to {\em collide} at time $t$ if $x_i(t)=x_j(t)$.

At any time, agents can be active or inactive.  
When activated, an agent $a_i$ 
performs  a sequence of operations called  \emph{Look-Compute-Move}:
 it activates the sensors  to obtain a snapshot (called {\em local view} )
 of the positions of the visible agents   expressed in its own coordinate system ({\em Look});
it then    executes an algorithm (the same for all agents) based on its  local view, which returns
 a destination point $x \in \mathbb{R}^{2}$ and a color $c \in C$  ({\em Compute}); it then
   sets  its own state variable  to $c$ and     moves  towards $x$ ({\em Move}).
  The movement  may be stopped by an adversary  before the
agent reaches its destination; the only constraint on the adversary is that, if interrupted 
before reaching its destination, a robot moves  at least a 
 minimum distance  $\delta>0$ (otherwise, no destination can ever be reached).

We consider two schedulers  for the activation of the agents:
{\em Semi-Synchronous} ({\em SSYNC}) and  {\em  Asynchronous} ({\em ASYNC}). 
In   {\em SSYNC},  the time is discrete; at each time instant $t$ (called {\em round}) 
 a subset of the agents is activated and performs its operational cycle instantaneously.
 The choice of the activation is done by an adversary, which however   activates each agent infinitely often.
In   {\em ASYNC},  there is no common notion of time; each agent is activated independently, and   each  Compute and Move operation can take an unpredictable (but bounded) amount of time, unknown to the agent. 

 At the beginning (time t=0), the agents start in an arbitrary configuration $\mathcal{C}_{0}$  occupying different positions, and they are \black\,  (the state variable of each one is set to a special symbol \blank\,).  The goal is for the agents to reach, in finite time,  an obstruction-free  configuration without ever colliding.
We call this problem {\sc Complete Visibility}. An algorithm is said to solve the problem if it always achieves complete visibility regardless of the choices of the adversary, and  from any initial configuration.

Let  $\mathcal{H}_{t}$   be  the convex hull defined by $\mathcal{C}_{t}$,
 let  $\partial \mathcal{H}_{t}=\mathcal{V}_{t} \cup \mathcal{B}_{t}$ denote the agents on the
 border of $\mathcal{H}_{t}$,
  where  $\mathcal{V}_{t}:\{v_{1},\ldots,v_{k}\} \subseteq \mathcal{A}$ 
 is the set of agents ( {\em corner-agents}) located at the  corners of   $\mathcal{H}_{t}$ 
 and  $\mathcal{B}_{r}:\{b_{1},\ldots,b_{l}\}$ is the set of those located   on the edges of $\mathcal{H}_{r}$
 ({\em  edge-agents}); let ${\cal I}_t$ be the set of agents that are interior of $\mathcal{H}_{t}$ ( {\em interior-agents}).
Let  $n_t = |\mathcal{V}_{t} |$ be the number of corners in $\mathcal{H}_{0}$. 
 Given an agent $a_{i} \in A$, we denote by  $\mathcal{H}_{t}[a_{i}]$  
   the convex hull of its local view $\mathcal{C}_{t}[a_{i}]$.
Let  $\mathcal{C}^{c}_{t}$ indicate the set of agents in $\mathcal{C}_{t}$ with color  $c$ at time $t$,    similarly we define  $\mathcal{H}^{c}_{t}[a_{i}]$ as the convex hull,     of $\mathcal{C}^{c}_{t}[a_i]$.  Analogously defined are the extensions of      $ \mathcal{V}_{t},\mathcal{B}_{t},\mathcal{I}_{t}$.
Given a configuration ${\cal C}$, we indicate by $SEC(\mathcal{C})$ the smallest enclosing circle containing  $\mathcal{C}$ (when no ambiguity arises we just use the term $SEC$). 
Given two points $x,y  \in R^{2}$ with $ {xy}$ we indicate the line that contains them, and we use the operator $\cap$ to indicate the intersection of lines and segments. Let  $d(x,y)$   
indicate the Euclidean distance between two points (or a segment and a point); moreover, given $x,y,z \in R^{2}$ we use $\angle xyz$ to indicate the angle with vertex $y$ and   sides $xy,yz$.  In the following, with an abuse of notation, when no ambiguity arises, we use $a_i$ to denote both the agent and its position.

\section{{  Complete Visibility}  in SSYNC\label{sec:SSYNC}}

In  this Section we  provide an Algorithm that reaches  Complete Visibility in  the semi-synchronous setting.
The algorithm is described assuming  $|\mathcal{V}_{0}| \geq 3$;  we will then show how the agents can easily move to reach this condition starting from a configuration with  $|\mathcal{V}_{0}| = 2$. 

Our  algorithm  works in two   phases: (1) {\em  Interior Depletion} (ID) and (2) {\em Edge Depletion} (ED). 
The purpose of the  Interior Depletion phase    is to reach a configuration ${\cal C}_{ID}$ 
 in which there are no  interior-agents. In this phase,  the interior-agents move towards an edge they perceive as  belonging to the border of the convex hull, and they position themselves between two corner-agents.  At the end of this phase, all agents are on $\partial \mathcal{H}_{0}$.
The goal of  the Edge Depletion phase is to have all agents  in  ${\cal B}_{ID}$   to
    move   so to reach  complete visibility. 

\subsection{Phase 1: Interior Depletion Phase \label{idp}} 

Initially all agents are \black.
The objective of this phase is to have  all agents on $\partial \mathcal{H}_{0}$, with the corner-agents colored  \red \, and the edge-agents colored  \brown.

Notice that corner (resp. edge) agents are able to recognize their   condition in spite of   possible obstructions. 
In fact, if a \black\, agent $a_i$  is activated at some round $r$, and it  sees that  ${\mathcal{C}}_{r}[a_i]$ contains a region of plane that is free of agents and wider than $180^{\circ}$,  then $a_i$ knows it is a corner
and  sets its variable $s_{i}$ to \red.  
A similar  rule is applied to edge-agents;  in this case, an edge-agent $a_i$  sets its variable $s_i$ to \brown \, if $C_{r}[a_i]$ contains a region of plane free of agents and wide exactly $180^{\circ}$ (see Coloring Case of Figure \ref{algorithm:id}). 

   In the ID phase, corner-agents color themselves \red, and no longer move, while edge-agents color themselves \brown.
Each interior-agent  $ a  $ moves to position itself on one of its nearest visible edges 
of $\partial \mathcal{H}_{0}$; note that an  edge  of $\partial \mathcal{H}_{0}$ can be recognized in $a$'s local view
once  it is occupied only by \brown\, and \red\, agents.
To prevent collisions, the   interior-agent moves   towards  the chosen   edge $e$ perpendicularly   if and only if 
it is one with  minimum distance to $e$ and its destination on $e$ is empty;  otherwise it does not move.
An edge-agent on the destination of an interior one, slightly moves to make room for the interior-agent.
The    {\sc Interior Depletion} algorithm is detailed in Figure \ref{algorithm:id}.

\begin{figure*}
\begin{framed}

{Algorithm  {\sc Interior Depletion}} (for the generic agent $a_i$ activated at round $r$)
\footnotesize
\begin{itemize}
 \item Coloring Case: if ($s_{i} =black$) then:
 \begin{itemize}
 \item If  ($a_{i}$ is a corner-agent in $\mathcal{H}_{r}[a_i])$) then $a_{i}$ sets $s_{i}=red$
  \item If ($a_{i}$ is an edge-agent in $\mathcal{H}_{r}[a_i])$) then $a_{i}$ sets $s_{i}=brown$
 \end{itemize}

 \item Interior Case: if ($a_{i}$ is interior in $\mathcal{H}_{r}[a_i])$ and $s_{i}=black$) then:
 \begin{itemize}
 \item $a_i$ uses its local view $\mathcal{C}_{r}[a_i]$ to determine  the edges 
 of  $  \partial \mathcal{H}_{r}[a_i]$.
  \item If ($\exists e \in     \partial \mathcal{H}_{r}[a_i]$ such that $\forall  a_j \in {\cal I}_r[a_i], d(a_j,e) \leq d(a_i,e)$) then 
  \begin{itemize}
  \item $a_{i}$ computes a point $x$ of $e$ such that $\overline{a_{i}x} \perp e$;
   if $x$ is empty, then $a_{i}$ moves   toward $x$
  \end{itemize}
 \end{itemize}

 \item Obstructing Edge Case: if ($s_i=brown$ ) then:

 \begin{itemize}
  \item Let  $e$ be the edge to which $a_{i}$ belongs; if ($\exists a_{j} \in {\mathcal{I}_{r}[a_{i}]} \wedge \overline{a_{j}a_{i}} \perp e$), then $a_{i}$ moves toward the nearest point $x \in e$ such that $\forall
  a_{k} \in  {\mathcal{I}_{r}}[a_{i}]$,  $\overline{a_{k}x} \not\perp e$. 
 
  \end{itemize}

\end{itemize}
\caption{\footnotesize Algorithm for the Interior Depletion Phase\label{algorithm:id}}
\end{framed}

\end{figure*}

It is easy to see that  at the end of this phase, all the  agents will be positioned on a convex hull.

 \begin{lemma}
\label{internalnodemove}
For any initial configuration $\mathcal{C}_{0}$ there exists a round $  r \in \mathbb{N^{+}}$ such that in $\mathcal{C}_{r}$ we have that $\mathcal{I}_{r}=\emptyset$; furthermore, this occurs without collisions.
\end{lemma}

\begin{theorem}\label{calIempty}
There is a round $r \in \mathbb{N^{+}}$ such that the  agents occupy   different positions
on $\mathcal{H}_{r}$.  Moreover, 
the corner-agents are \red, and the edge-agents are  \brown.
\end{theorem}

\subsection{Phase 2: Edge Depletion -ED}

 The purpose of the ED phase is to move the edge-agents out of the current convex hull  to reach a final configuration whose
convex hull includes  $\mathcal{H}_{0}$ and all agents are on the corners, thus achieving complete visibility.

\begin{figure}[H]

\hspace{-0.5cm}
        \begin{subfigure}[b]{0.53\textwidth}
                \includegraphics[width=\textwidth]{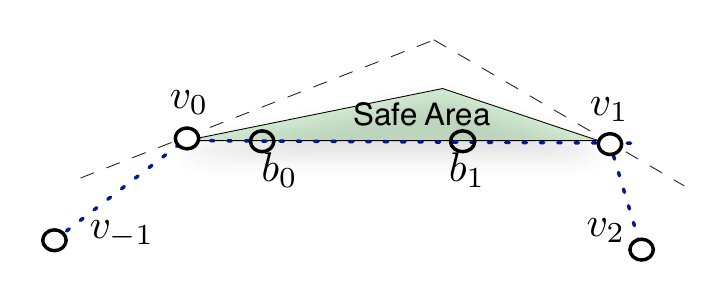}
                
                 \caption{ \scriptsize Safe Area of edge $\overline{v_{0}v_{1}}$: an agent moving inside the safe area cannot create collinearity with agents on the neighboring edges}
                \label{fig:safe}
        \end{subfigure}%
        ~ 
        \begin{subfigure}[b]{0.53\textwidth}
                \includegraphics[width=\textwidth]{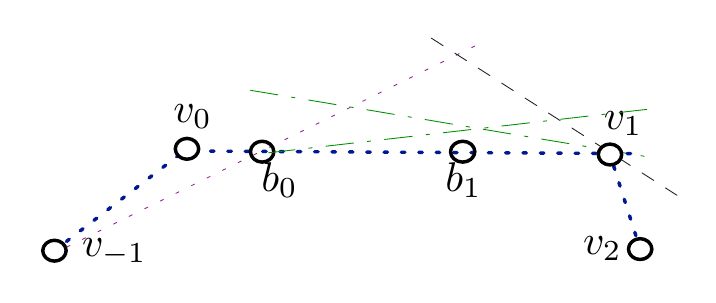}
                \caption{\scriptsize Approximation of the safe area computed by agent $b_{1}$ using as reference the two lines $v_{1}v_{2}$ and $v_{-1}b_{0}$. This approximation is entirely contained in the real safe area}
                \label{fig:safeapprox}
        \end{subfigure}

\hspace{-0.5cm}
             \begin{subfigure}[b]{0.53\textwidth}
                \includegraphics[width=\textwidth]{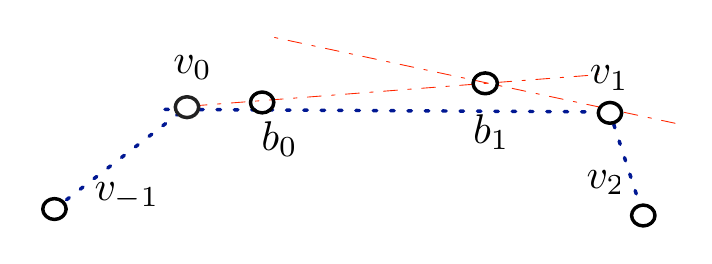}
                \caption{\scriptsize Creation of a new edge, due to non-rigid movements or to different approximations of the safe area, agent $b_{1}$ could move  making agent $b_{0}$ interior, this condition is adjusted by letting $b_{1}$ move towards $v_{1}$}
                \label{fig:yellowmove}
        \end{subfigure}
        ~   
                \begin{subfigure}[b]{0.53\textwidth}
                \includegraphics[width=\textwidth]{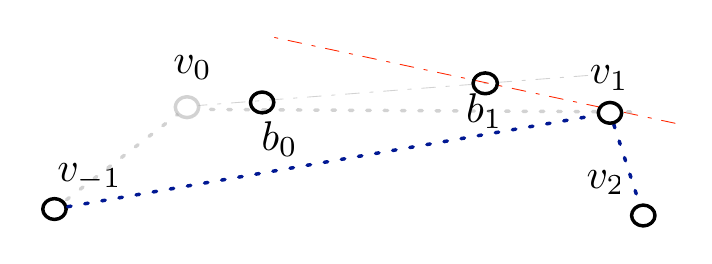}
                \caption{\scriptsize $b_{1}$ could move in such a way to become collinear to $v_{0}b_{0}$, $b_{0}$ signals this condition by changing its color}
                \label{fig:oran}
        \end{subfigure}
        \caption{Edge Depletion Phase         \label{fig:ed}}

\end{figure}

The algorithm makes an edge-agent move  from   its edge $e=\overline{v_{0}v_{1}}$   to a point 
 out of the current  convex hull, but    within a {\em safe zone}.
 Safe zones are calculated so to guarantee that \red \,   agents never cease to be located on corners of  the current convex hull,
   in spite of the movement of the edge-agents.
More precisely, the safe zone $S(e)$ of $e$  consists of  the portion of plane  outside the current convex hull,
 such that $\forall x \in S(e)$ we have $\angle xv_{0}v_{1} < \frac{180^{\circ}- \angle v_{-1}v_{0}v_{1}}{4}$ and $\angle v_{0}v_{1}x < \frac{180^{\circ}- \angle v_{0}v_{1}v_{2}}{4}$  (see Figure \ref{fig:safe}). 
 
Note that, due to the mutual obstructions that lead to different local views,     edge-agents cannot always compute $S(e)$ exactly (see Figure \ref{fig:safeapprox}). In fact, only when  there is a single edge-agent  between the two \red \,  corner-agents on $e$, the computation of $S(e)$ is exact; in any case,   we can show that the safe area $S'(e)$ computed by an agent   is $S'(e) \subseteq S(e)$ and thus still safe.

The migration of edge-agents and their transformation into corner-agents occurs in steps: in fact, if the  edge $e$ contains more than one edge-agent, our algorithm makes 
them move in turns, starting from the   two agents $b_{1}$ and $b_{0}$  that are immediate neighbors of the corners 
$v_{1}$ and $v_{0}$, respectively. Only once they are out of the convex hull and they are corner of a new edge $e'$,   other agents on $e$ will follow, always moving perpendicularly  to $e'$. Careful changes of colors are required to coordinate this process. In fact, once the first pair is in position, the two agents will become \blue \, to signal   the other \brown\, agents on $e$   that it is their turn to move out; they will  set their color  to \red\,  only when there is no interior-agent in the space delimited by $e'$ and $e$. Once \red, their color will never change until completion.

Due to the different estimations of $S$, to semi-synchronicity, and  to the unpredictable distance traversed by an agent (possibly stopped before destination), a variety of situations could disrupt this ideal behaviour.
In particular, it could happen that only one of the two agents, say $b_1$, moves  while the other stays still, or that
 $b_1$ moves further from $e$ than $b_0$. In both cases this leads to a configuration in which $b_{0}$ becomes a
 interior or   edge-agent.  
  This problem is however  adjusted by $b_{1}$ that, when noticing the situation,
 moves towards $v_{1}$ until $b_{0}$ becomes a corner in $\mathcal{H}[b_{1}]$. 
A further complication is  that  $b_{1}$ might wrongly perceive  $b_{0}$ as a corner and thus decide not to move;  this occurs if $v_{0} b_{0}$  happens to be collinear with $b_{1}$ obstructing visibility;  such a case is however detected by $b_{0}$ itself, which uses a different color (\orange)   to signal that $b_{1}$ has to move further towards $v_{1}$  to transform  $b_{0}$ into a corner (see Figure \ref{fig:oran}). 


The detailed algorithm for the ED phase is reported in Figure \ref{algorithm:ed}.  

 \begin{figure}
\begin{framed}
 Algorithm  {\sc Edge Depletion}\\ 
  \footnotesize
 For agent $a_i$ activated at round $r$; to be executed  if and only if $\nexists (black,a_j) \in \mathcal{C}_{r}[a_i]$.\\

 -  Execute {\sc Compute  Order} and appropriate case from the list below.

\begin{itemize}

 \item {\sc Brown Edge Case}:  $a_i$ belongs to an edge $e$ of $\mathcal{C}_{r}[a_i]$ and $s_i=brown$. 
 
 If $a_i$ is the only agent on $e$ then 
 
 \begin{itemize}
 \item $a_i$  computes the  angles $\alpha = 180^\circ  - \angle v_{-1}v_{0}a_{i}$, $\beta= 180^\circ -\angle a_{i}v_{1}v_{2}$,  and $\gamma=min(\frac{\alpha}{4},\frac{\beta}{4})$; it then computes a point $x$ such that $\angle xv_{1}a_{i}  < \gamma$ and $\angle xv_{0}a_{i} < \gamma$. 

  \item $a_i$ sets $s_i=yellow$.
  \item $a_i$ moves perpendicularly to $e$ with destination $x$.
   \end{itemize}
   If $a_i$ is not the only non-\red \ agent on $e$ and one of its neighbors on $e$ is  \red\ (by routine {\sc ComputeOrder},  this agent is $v_{1}$)  then: let $b$ be its other neighbor;
     \begin{itemize}
     \item $a_i$ computes the two angles $\alpha=180^\circ  - \angle a_{i}v_{1}v_{2}$,  $\beta=180^{\circ} - \angle v_{-1}ba_{i}$, and $\gamma=min(\frac{\alpha}{4},\frac{\beta}{4})$;  it then computes a point $x$ such that
     $\angle xv_{1}a_{i}  < \gamma \,\, \wedge \,\, \angle xba_{i}  < \gamma$.
  \item $a_i$ sets $s_i=yellow$.
  \item $a_i$  moves perpendicularly to $e$ with destination  $x$. 
   \end{itemize}
   
   \item {\sc Yellow Case}: $s_{i}=yellow$.
   \begin{itemize}
                      \item if there is another \yellow\ or \blue\ agent $a_j$ with $e^{a_{i}}=e^{a_{j}}$ then
               \begin{itemize}
             	\item if $a_{i}v_{1} \cap a_{j}v_{{0}} \not\in (\overline{a_{i}v_{1}} \cup \overline{a_jv_{{0}}})$ then $a_i$ sets $s_i=blue$
		\item if $a_{i}v_{1} \cap a_{j}v_{{0}} \in \overline{a_{i}v_{1}} $ then $a_i$ moves towards $v_{1}$ along $\overline{a_{i}v_{1}} $ of $\frac{d(a_i,v_{1})}{2}$
             	\item  if $ a_{j} \in a_{i}v_{1}$ then $a_i$ sets $s_i=orange$
       		\end{itemize}
        \item else if $\nexists (s_j,a_j) \in {\mathcal{C}_{r}[a_i]}$ with $a_j \ne a_i$ and $e^{a_i}=e^{a_j}$ and $\nexists (s_j,a_j) \in {\mathcal{C}_{r}[a_i]} 	\cap e^{a_i}$ then 
         \begin{itemize}
         	\item $a_i$ set $s_i=red$
         \end{itemize}
    \end{itemize}
 \item {\sc Orange Case:} $s_{i}=orange$.
   \begin{itemize}
      
        \item if there is another blue agent $a_j$ with $e^{a_{i}}=e^{a_{j}}$ then
               \begin{itemize}
                     	\item  if $ a_{j} \not\in a_{i}v_{1}$ then $a_i$ sets $s_i=blue$

       		\end{itemize}

    \end{itemize}
    
     \item {\sc Blue Case:} $s_{i}=blue$ and $a_{i} \in e$ with $e$ edge of ${\mathcal{C}_{r}}[a_i]$.
     \begin{itemize}
     	 
	 \item  if there is another orange agent $a_j$ with $e^{a_{i}}=e^{a_{j}}$ then
          \begin{itemize}
	        \item $a_i$ moves  along $\overline{a_{i}v_{1}} $   in direction of  $v_{1}$   towards the point  at distance
	            $\frac{d(a_i,v_{1})}{2}$
          \end{itemize}
          \item else if  $\nexists (brown,a_j) \in{\mathcal{C}_r}[a_i] $     such that $a_j$ could move to $e$ then
          	\begin{itemize}
			\item $a_i$ sets $s_i=red$
		\end{itemize}
     \end{itemize}
      \item {\sc Brown Interior Case:} $a_i$ is such that $s_i = brown$ and $a_i \in {\mathcal{I}_{r}}[a_i]$.
 \begin{itemize}
 \item if there exists and edge $e'=\overline{a_{x}a_{y}}$ with $s_{x}=s_y=blue$ and $a_{i}$ could move perpendicularly towards $e'$ without crossing any segment delimited by two \red\, agents,  then $a_{i}$ moves towards $e'$.  
 \item if $a_i \in e= \overline{a_{0}a_{1}}$ with  $e \in \partial {\mathcal{H}^{\{red,brown\}}_{r}}[a_i]$ and $\exists x \in \mathbb{R}^{2} $  such that $a_{0}a_{i} \perp \overline{xa_{i}}$ and $\nexists a_{j} \in \angle a_{0}a_{i}x$ or $ \nexists a_{j} \in \angle a_{1}a_{i}x$ then $a_{i}$ executes the second subcase of  the {\sc Brown Edge Case}.
  \end{itemize} 
     
 \item {\sc Corner Case}: $a_i$ is a corner of ${\mathcal{C}_{r}}[a_i]$ and $s_i=red$.
 \begin{itemize}
 \item $a_i$ can check   local termination and the global termination
 \begin{itemize}
 \item $a_i$ locally terminates when $s_i=red$
 \item $a_i$ detects the global termination of ED phase when $\nexists (s_j,a_j )\in {\mathcal{C}_{r}}[a_i]$ with $s_j \neq red$
 \end{itemize}
 \end{itemize}

\end{itemize}
\caption{\footnotesize Edge Depletion Phase algorithm \label{algorithm:ed}}

\end{framed}
\end{figure}

\begin{figure}
\begin{framed}
Procedure {\sc Compute Order}
 \footnotesize 
\begin{itemize}
 
 \item  if  $a_i$ belongs to an edge $e$ of $\mathcal{H}_{r}[a_i]$ and $s_i=brown$,
  it orders  the \red\ agents in its local view in a circular order, starting from the closest, 
  $(v_1,v_2,\dots,v_0)$.
 
    \item  if $s_{i} \in \{orange,blue,yellow\}$, then 
 $a_{i}$ determines which of its current neighbors was $v_1$ in its previous computation and 
                      the edge $e^{a_i} = \overline{v_1v_{0}}$ to which it belonged:
                      \begin{itemize}
                      \item $a_{i}$ computes the nearest edge $e=\{u,v\} \in \mathcal{H}^{red}_{r}[a_{i}]$
                      \item $a_{i}$ computes the point $x \in \mathbb{R}^{2}$ such that is $uv \perp a_{i}x$
                      \item $a_{i}$ sets $v_{1}=u,v_{0}=v$ if $\nexists a_{j} \in \angle uxa_{i}$ otherwise it sets $v_{1}=v,v_{0}=u$.
                      \item $a_{i}$ sets $e^{a_i}=\overline{v_{1}v_{0}}$
                      \end{itemize}

 \end{itemize} 
 \end{framed}
\end{figure}

\begin{figure}
\footnotesize
\begin{center}
    \begin{tabular}{| l |p{8cm}|  l   |}
    \hline
  {\bf   Color} &  {\bf   Meaning}  & {\bf Transition to:}\\ \hline
    \hline
    $Black$ &   initial color of all agents & $\{Red,Brown\}$ \\ \hline
    $Brown$  &   agents on   edges or having  to move to a new edge of  ${\cal H}$ & $Yellow$ \\ \hline
    $Yellow$  &     agents   moving out of   ${\cal H}$ to form a new edge  & $\{Blue,Orange,Red\}$  \\ \hline
    $Orange $   &  agents needing to be transformed into corners & $Blue$  \\ \hline
     $Blue $ &     corner-agent now forming a new edge $e$, waiting for other agents to move to $e$  & $Red$  \\ \hline
             $Red$     &        a stable corner-agent   & $-$ \\ \hline
    \end{tabular}
\end{center}
\caption{\footnotesize Colors used in the  {\sc Complete Visibility} algorithm. \label{colors}}
\end{figure}

 \subsection{The case of   $|\mathcal{V}_{0}| =2$}

The   strategy of the previous Section works for $|\mathcal{V}_{0}| >  2$.  It is however  simple to have the agent move to reach such a condition from  $|\mathcal{V}_{0}| = 2$, as described below.

When $|\mathcal{V}_{0}| = 2$ the agents are necessarily disposed forming a line and  $|\mathcal{A}| \geq 2$.
First notice that  an agent $a$ can  detect that the configuration is a line, and whether it is an extremity (i.e., it sees only one other agents $a'$), or an internal agent  (i.e., it is between two collinear agents).
If     $a$  is internal, it does not move; if it is an extreme,      $a$   sets its color  to \red \, and   moves perpendicular to the segment $\overline{a'a}$. This means that,   as soon as at least one of the extremes is activated, it will move (or they will move) creating  a configuration with $|{\cal V}| > 2$.  At that point,  the algorithm previously described  is applied.

\subsubsection{Correctness of the ED phase.}

With the following lemma we show that the global  absence of interior-agents with respect to the initial convex hull, can be locally detected  by each agent.

\begin{lemma}\label{lemma:edstart} 
Given an agent $a_i \in \mathcal{A}$ with $s_i \in \{red,brown\}$ and a round $r \in \mathbb{N}^{+}$,
 if  $\nexists (black,a_j) \in \mathcal{C}_{r}[a_i]$ then $\mathcal{C}_{r}$ does not contain interior-agents with respect to $\mathcal{H}_{0}$.
\end{lemma}
\begin{proofL}
By contradiction,  assume that  $\nexists (black,a_j) \in \mathcal{C}_{r}[a_i]$ but there exists at least an interior-agent $a$ with respect to  $\mathcal{H}_{0}$. 
By  the rules of the ID phase, agent  $a$ cannot change its color from $black$ to another
because it can detect it is neither a corner nor a border.
   Thus, $a$ is not in $\mathcal{C}_{r}[a_i]$ because $\mathcal{C}_{r}[a_i]$, by assumption,  does not contain \black\, agents. Thus, it must exist an agent $a_k$ that has color different from $black$ and  $a_k \in  \overline{a_ia}$. But since $a$ is interior then also $a_k$ is interior, and so $s_k=black$.
\renewcommand{\toto}{lemma:edstart}
\end{proofL}

We now show that the safe area $S'(e)$ computed by an edge-agent on $e$  is such that $S'(e) \subseteq S(e)$ and thus its movement is still safe (it does not transform a \red\, corner into an interior or edge-agent).

\begin{lemma}\label{lemma:safeapprox} 
Given a configuration $\mathcal{C}_{r}$ and an edge $e=\overline{v_{0}v_{1}}$ of $ \mathcal{H}_{r}$, if an agent  $a_{j} \in e$ moves from $e$,  it moves inside the safe zone $S(e)$  
\end{lemma}
\begin{proofL}
The case when there is a single edge-agent $b \in e$ is trivial because  $b$ can compute exactly $S(e)$. 
Consider now the case when there are   two or more edge-agents on $e$; among those, let $b_{0}$ and $b_{1}$  be the two that
are  neighbors of $v_{0},v_{1}$. 
Those agents move only when executing the  Brown Edge Case or Brown Interior Case.
Let us consider the movement of the first that is activated,  say $b_{1}$. Agent $b_{1}$ has two neighbors on $e$: a \brown\, neighbor   $b$ and the \red\,  corner $v_{1}$. Agent   $b_{1}$   orders   the corners in its view from $v_{1}$ to $v_{last}$, according to its local notion of clockwise, where  $v_{last}$ is the last corner before $b$, i.e. $v_{-1}$ in Figure \ref{fig:safeapprox}. 
Following the rules of the algorithm, $b_1$   computes:
$\alpha = 180^\circ  - \angle v_{last}bb_{1}$, $\beta = 180^\circ -\angle b_{1}v_{1}v_{2}$, and    $\gamma=min(\frac{\alpha}{4},\frac{\beta}{4})$. Angle  $\angle v_{last}bb_{1}$ is an upper bound on $\angle v_{last}v_{0}b_{1}$, otherwise we could get a contradiction since $v_{last}b$ and $v_{last}v_{0}$ will intersect in two points: one is $v_{last}$ and the other one is after the intersection of $v_{last}v_{0}$ and $v_{0}v_{1}$, that is impossible. Thus, $\alpha$ is a lower bound on the  angle that a single agent would compute on $e$, which implies that $b_{1}$ will move inside $S(e)$. The same holds for $b_{0}$.  Notice that, given two points $x$ and $y$ inside the safe zone, any point $z \in \overline{xy}$ is still inside the safe zone, thus any agent  that   moves on the lines connecting two agents inside $S(e)$ will still be in $S(e)$, completing  the proof. 
 \renewcommand{\toto}{lemma:safeapprox}
\end{proofL}

The next lemma shows that the moves of our algorithm cannot transform any \red\,  corner-agents into an interior-agent.  

\begin{lemma}\label{lemma:nodecreasing} 
Consider a corner-agent $v_1$ of $ \mathcal{H}_{r'}$ with $s_{1}=red$,
 we have that $\forall r \in \mathbb{N}^{+}$ with $r > r'$, 
 $v_1$ is also a \red\, corner-agent of $ \mathcal{H}_{r}$.
\end{lemma}
\begin{proofL}
 It is easy to see that during the ID phase we have that $\mathcal{H}_{r}=\mathcal{H}_{0}$ since the interior-agents will never trespass the edges of $\mathcal{H}_{0}$, so the hypothesis holds.
  We have to show that the same holds during the ED phase. We have that $v_1$ never moves after it sets $s_1=red$ so if $v_1$ is a corner it cannot become interior as a consequence of its own move. 
Consider the two   edges adjacent to $v_{1}$:  $e_{1}=\overline{v_{0}v_{1}}$ and $e_{2}=\overline{v_{1}v_{2}}$. Assume,  by contradiction, that there  exists a round $r$ in which the moves of a set $X$ of agents on these two edges is such that $v_1$ is a corner-agent in  $\mathcal{H}_{r-1}$ but not in $\mathcal{H}_{r}$. 
From Lemma \ref{lemma:safeapprox} we have that agents in $X$ move to points inside the safe zones $S(e_1)$ and $S(e_2)$ of $e_{1},e_{2}$. Let us consider two points $x \in S(e_1)$ and $y \in S(e_2)$, such that agents on them will make $v_{1}$ interior. If $v_{1}$ is interior in  $\mathcal{H}_{r}$, we have that  $\angle xv_{1}y > 180^{\circ}$. It is easy to see that $\angle v_{0}v_{1}x  < \gamma$ (see Brown Edge Case and Brown Interior Case of Figure \ref{algorithm:ed}) and   that $\gamma \leq \frac{180^{\circ} - \angle v_{0}v_{1}v_{2}}{4}$,  since $\gamma=min(\frac{\alpha}{4},\frac{\beta}{4})$,
 and that at least one of the two among $\beta,\alpha$ is a lower bound on $180^{\circ}-\angle v_{0}v_{1}v_{2}$.  
The same holds for $y$,  so we have $\angle v_{2}v_{1}y \leq \frac{180^{\circ} - \angle v_{0}v_{1}v_{2}}{4}$. 
Thus, we have $ \angle v_{0}v_{1}x +\angle v_{2}v_{1}y+\angle v_{0}v_{1}v_{2} < 180^{\circ} $ and  then  $\angle xv_{1}y < 180^{\circ}$, which is a contradiction. So, $v_{1}$ cannot be interior in $\mathcal{H}_{r}$. The same arguments hold  if at round $r-1$ we consider a set of agents $X$ on two edges $e',e''$ that are not adjacent to $v_{1}$; this is easy to see since,
 given $x \in S(e')$ and $y \in S(e'')$ we have $\angle{xv_{1}y} \leq \angle{v_{0}v_{1}v_{2}} < 180^{\circ}$, which  is another contradiction to the hypothesis of $v_{1}$ being  interior in $\mathcal{H}_{r}$.   
  \renewcommand{\toto}{lemma:nodecreasing}
\end{proofL}
 
In the next   sequence of lemmas, we show that, given an edge $e$ in a configuration $\mathcal{C}$ of the ED phase, all edge-agents in $e$ will eventually became \red \, corners.

\begin{lemma}\label{lemma:y} 
Given a configuration $\mathcal{C}_{r}$ and an edge $e$ of $\mathcal{H}_{r}$ with 
a single  brown  agent $b$ on $e$, eventually $b$ will be a \red\,  corner.  
 \end{lemma}
\begin{proofL}
Since   \red \, corners never move and no interior-agents can be moving on $e$,
while inactive, agent $b$  maintains its single position inside $e$.
When activated at some round $r'$, agent $b$    executes the Brown Edge Case with a single agent. 
Thus $b$   switches color to \yellow\, and it  moves   perpendicularly to $e$ of at least $min(d(v_h,x),\delta)$. At round $r'+1$,
 $b$ is a corner-agent of $\mathcal{H}_{r'+1}$; in the next activation, after  executing the Yellow Case code,   
 $b$     becomes  \red.
\renewcommand{\toto}{lemma:y}
\end{proofL}

\begin{lemma}\label{lemma:ye} Given a configuration $\mathcal{C}_{r}$ and an edge $e$ of $\mathcal{H}_{r}$ with  exactly two brown agents $b_{0},b_{1}$ on $e$,
eventually  they  will  set  their state variable  to $yellow$ and they will move outside $e$. 
\end{lemma}
\begin{proofL}
Let $b_{1}$ be the first to be activated   at some round $r' \geq r$. At that time, $b_{1}$   switches its color  to \yellow\,  and it   moves perpendicularly to $e$ (see Brown Edge Case).   Agent $b_{0}$ will do the same, no matter   if it is activated in round  $r'$ or in some successive rounds (see Brown Edge Case and Brown Interior Case).   
\renewcommand{\toto}{lemma:ye}
\end{proofL}

\begin{lemma}\label{lemma:yv} 
Given a configuration $\mathcal{C}$, 
any    agent   $b_{1}$ with $s_{1}=yellow$  eventually  becomes
 corner and will  sets its state variable to $red$. \end{lemma}
\begin{proofL}
If $b_{1}$  is \yellow \, then $a_{1}$ has moved from an edge $e=\overline{v_0v_1}$.  
If $b_{1}$ was not the only agent on $e$ that could move, 
then there is (or there will be)  another \yellow\, agent $b_{0}$ 
moving from $e$. By construction, $b_{1}$ waits until it sees the other \yellow \, agent $b_{0}$ (see Yellow Case).
 If both $b_{1}$ and $b_{0}$ realize to be corners of the current  convex hull, then they eventually     set their  color to \blue\, and then to  \red, thus the lemma is proved. 
 However,   due to the non-rigidity or the different local views of $b_{1}$ and $b_{0}$, the pathological case of Figure \ref{fig:yellowmove} may arise where one of the two, say  $b_{0}$, becomes an interior-agent. 
 This case is adjusted by the Yellow Case rule: each time $a_{1}$ is activated, it will move towards $v_{1}$ until   a round $r''$ is reached when   $b_{0}$ is not   interior anymore in ${\mathcal{C}_{r''}}[b_{1}]$. Note that, since $b_{1}$ moves always half of the distance $d(b_{1},v_{1})$, and the number of rounds  until the next activation of  $b_0$ is finite,
  we have that $b_1$ will never touch $v_{1}$.  Two possible sub-cases may happen  at round  $r''$:
  $(i)$  $b_{1}v_{1} \cap b_{0}v_{{0}} \not\in (\overline{b_{0}v_{0}} \cup \overline{b_{1}v_{1}})$:  in this case, in the subsequent  activations, $b_{1}$ and $b_{0}$ will set their colors to \blue;
  $ (ii)$  $b_{1} \in b_{0}v_{0}$:  this might not be detected by the local view of $b_{1}$, but it is detected by $b_{0}$ that   sets its color  to \orange\,; in the next activations $b_{1}$ will move so to transform  $b_{0}$ into a corner  and, after this move, an activation of $b_{0}$ will set $s_{0}=blue$. So, in  both sub-cases we eventually reach   a configuration in which $b_{1}$ and $b_{0}$ are \blue \,   corner-agents. In the subsequent activations, they will set their color to \red\,, proving the lemma.  
\renewcommand{\toto}{lemma:yv}
\end{proofL}

\begin{lemma}\label{lemma:by}Given a configuration $\mathcal{C}$, let   $e=\overline{v_{0}v_{1}}$  be an edge 
with $q>2$ edge-agents on it. Eventually all these agents will become corners and set their color to \red.\end{lemma}
\begin{proofL}
The two edge-agents $b_{0},b_{1} \in e$  that are neighbors of \red\, corners,   execute the same code described in the previous lemma.  So, they  wil reach a configuration $\mathcal{C}_{r'}$ in which $b_{0}$ and $b_{1}$ are \blue\,   corner-agents. In this case, they wait until all the agents on $e$ move on the segment $\overline{b_{0}b_{1}}$; then, they   set their color to \red\,  (see the rule 3 of Blue Case).  It is straightforward to see that each remaining agent on $e$ will move now towards this new edge without colliding, since all movements to the same edge are  on parallels trajectories. It follows that,  in  finite time, a new edge $e'$ is formed with $q-2$ agents. Iterating the reasoning we will end up in a case where the number of edge-agents on the same edge is at most 2, hence,  by Lemmas \ref{lemma:y}-\ref{lemma:yv},
  the lemma follows. 
\renewcommand{\toto}{lemma:by}
\end{proofL}

 \begin{theorem}  
The  {\sc Complete Visibility} problem is solvable in SSYNC by a team of oblivious, obstructable agents, using five colors without creating any collision.
\label{t2}
\end{theorem}
\begin{proofT} From Theorem \ref{calIempty} we have that from any configuration $\mathcal{C}_{0}$ we reach a configuration $\mathcal{C}_{ID}$ where $\mathcal{I}_{ID}=\emptyset$. This is locally detected by  agents (see Lemma \ref{lemma:edstart}), that start executing the ED phase. 
By  Lemma  \ref{lemma:nodecreasing} we have that the number of \red\, corners is not decreasing during the execution of the algorithm. From Lemmas \ref{lemma:y}-\ref{lemma:by} we have that eventually each edge-agent $a$ of $\mathcal{H}_{ID}$ will became a \red\, corner. So we will reach a configuration $\mathcal{C}_{final}$ in which all agents are corner of $\mathcal{H}_{final}$, thus,  they cannot obstruct each other.
Moreover, It is easy to see that each agent is able to detect not only   local termination, when it sets its color to \red, but also global termination of ED phase, and thus of the algorithm, when each agent in its local view is \red.
\renewcommand{\toto}{t2}
\end{proofT}

\section{{  Complete Visibility}  in ASYNC \label{async}}

In this section we consider the  asynchronous model (ASYNC), where there is no common notion of time or rounds,
there are no assumptions on time, on activation, on synchronization;  moreover, each  Compute and Move operation 
and inactivity may take ant unpredictable  (but finite) amount of time, unknown to the agent.  As a consequence, agents can be seen while
moving, and their computations and movements may be based on obsolete information.

\paragraph{\bf Asynchronous Interior Depletion phase.}

The {\sc Interior Depletion}  algorithm  of Sec. \ref{idp} works also in ASYNC without modifications. We only need to show that  the asynchronous behaviour  of the  agents, and in particular the asynchronous assignment of colors,
 cannot induce a collision among interior-agents.  
Since agents always move perpendicularly to the closest edge, it is easy to see that this does not happen  and thus   Lemmas \ref{internalnodemove}   and Theorem \ref{calIempty} hold also  in the asynchronous case.

\paragraph{\bf Asynchronous Edge Depletion phase.}

The Edge Depletion phase has to be modified for ASYNC. 
To see why the {\sc Edge Depletion} algorithm would not work,  consider, for example,  the Yellow Case in Algorithm \ref{algorithm:ed}:  it is possible that a moving \yellow\, agent is seen by another \yellow\, agent, this could lead to scenarios in which an agent assumes   color \red\, while it is on the edge of the convex hull and not on a corner. 

The source of  inconsistencies is the fact that agents can be seen while in transit.
 To prevent this problem we use new colors (\yellow\_\moving\, and  \blue\_\moving\,) to signal 
 that the agents are in transit;
 those agents     will take 
 color \yellow\, (resp. \blue) once as the movement is completed.
Using these intermediate colors, we can    simulate  the ED phase  of the previous Section 
(for   $|\mathcal{V}_{0}|>  2$).
 
 More precisely, in the {\em Edge Depletion} algorithm  of Figure \ref{algorithm:ed},
  instead of becoming \yellow, a \brown\, agent becomes  \yellow\_\moving, turning \yellow\
 at the next activation. Similarly,  instead of becoming \blue, a \yellow\, agent becomes  \blue\_\moving, 
 turning \blue\, only when seeing that the ``companion" agent is   \blue\_\moving\, or \blue.

It is not difficult  to see that, with these additional colors, since agents will always move inside the safe zones of $\mathcal{H}$, the  validity of Lemmas \ref{lemma:nodecreasing}, \ref{lemma:yv}-\ref{lemma:by} holds also in ASYNC.

\paragraph{\bf The case of   $|\mathcal{V}_{0}| = 2$.}

When the agents initially form a line, the algorithm described for SSYNC where the agents first move 
to a configuration $|\mathcal{V}_{0}| >  2$, and then apply the general Algorithm, would not work.
Consider, for example, the following scenario:   both   extreme agents compute  and their destination is in   opposite direction,  but only one of them actually moves. At this point, the   agents on the line set their color to \red\, or \brown\,, but they will became interior-agents as soon as the slower  extreme agent  moves from the line towards its destination, thus  changing the convex hull.

\begin{figure}[H]

             \begin{subfigure}[b]{0.5\textwidth}
                \includegraphics[scale=0.5]{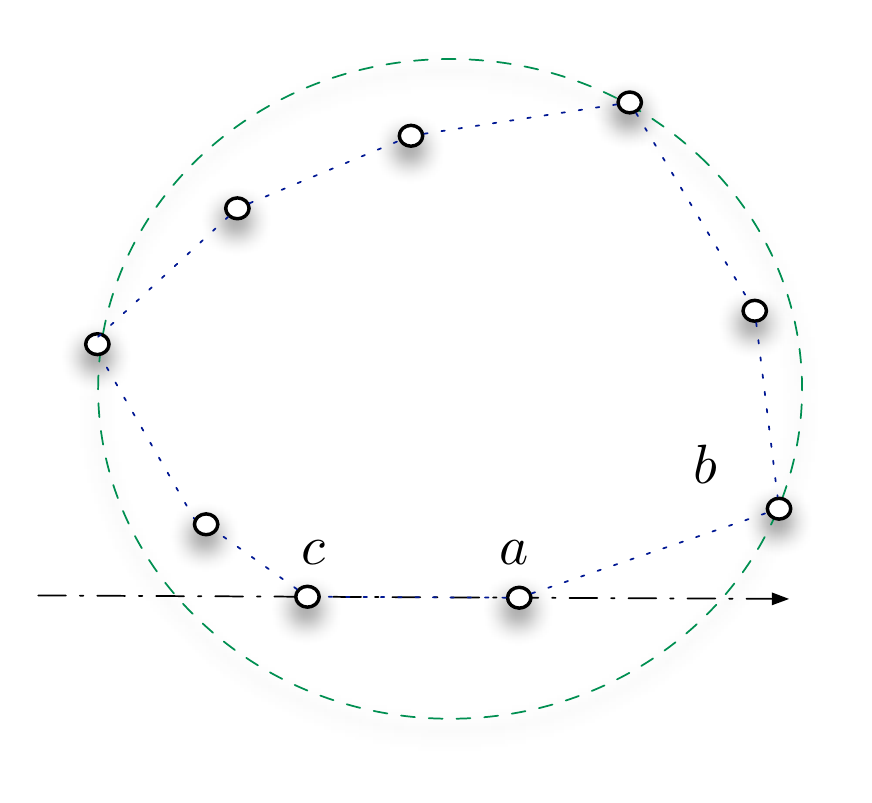}
                \caption{\scriptsize {\sc Circle Formation}: Agent $a$ is neighbor of an agent $b$ on   $SEC$, so it   moves on   line $ca$ in direction of   $SEC$.   During this movement, the  corner-agents of the convex hull are not modified, and 
  visibility with the other agents is preserved. \label{rule1}}
                \label{rule1}
        \end{subfigure}
        ~
                \begin{subfigure}[b]{0.5\textwidth}
                \includegraphics[scale=0.55]{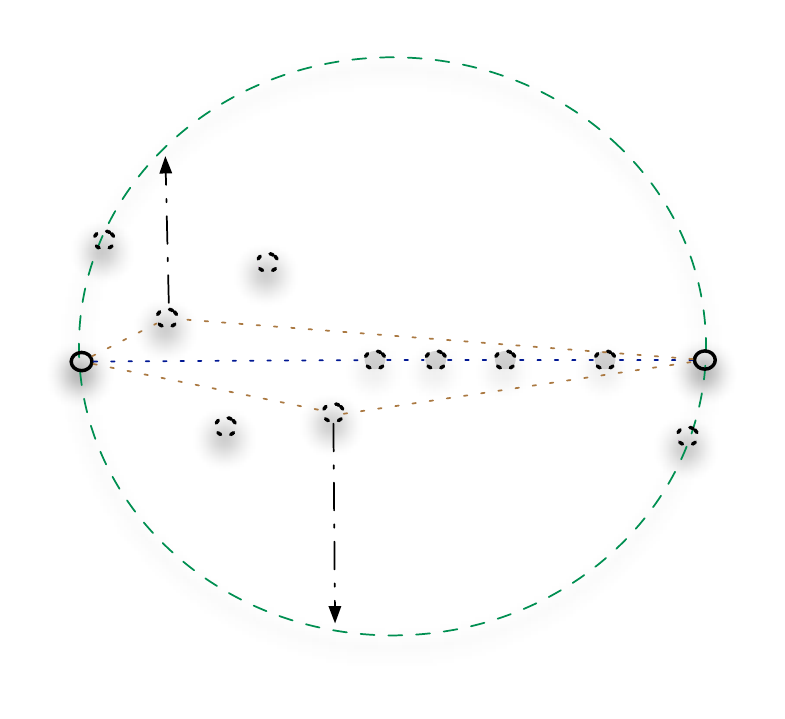}
                \caption{\scriptsize {\sc Asynch formation for $|\mathcal{V}_{0}|=2$}: the two extreme agents signal the line configuration with   color {\em line-extreme}, the other agents move perpendicularly to them until they reach the $SEC$ whose  diameter   is defined by the extreme agents. }
                \label{fig:collinear}
        \end{subfigure}
        \\
ption{Edge Depletion Phase         \label{fig:ed}}

\end{figure}

The  idea is to use a completely different algorithm in ASYNC when the initial configuration is a line (refer to   Figure \ref{fig:collinear}). 
Two additional  special    colors ({\em line-extreme} and {\em line-moving}) are used. 
The color
{\em line-extreme}  is taken by the two agents $a_1$ and $a_2$  located at the  extreme points of the line, $x_1$ and $x_n$, when activated; this color is used
to  acknowledge the line  condition, and to define the smallest enclosing circle $SEC$ with diameter $\overline{x_1,x_n}$. Notice that, due  to obstructed visibility, the diameter, and thus $SEC$, is unknown to the agents.
The  two extreme agents will never move. 

The general  strategy is to have the other agents move to points on  $SEC$. 
 First notice that an agent $a$ can detect that the configuration is a line, either by geometric conditions   (i.e., it sees only one or two collinear  agents), or by the special color of  some visible agents ({\em line-extreme} or {\em line-moving}).
  If an uncolred agent  $a$ located in $x$  sees a {\em line-extreme} agent (say $a_1$),
 then $a$  changes its color to {\em line-moving} and   it moves   perpendicularly to  $x x_1$ toward the perimeter of the circle whose diameter is identified by $a_1$ and the closest agent $b \neq a_1$ on the line $ x x_1$ (note that there must be at least one, possibly the other extreme).
 A {\em line moving} agent follows similar rules; if it can detect $SEC$ (e.g., it sees two {\em line-extreme}) it continues its perpendicular move towards it.
   Otherwise, it does not move.
 It can be shown that, at any time, there is   at least one agent  that, if activated, can  move. 
  A non extreme  
   agent   switches its color to \red  \,  when it sees only agents on the SEC;
an extreme agents switches its  color to \red\, when it sees only \red \, or {\em line-extreme} agents. 
   
 It is not difficult  to see that this set of rules will allow  the agents to   reach   $SEC$ in finite time becoming \red, and thus to solve the {\sc Complete Visibility} problem.

\begin{theorem}  
The  {\sc Complete Visibility} problem is solvable in ASYNC by a team of oblivious, 
obstructable agents, using eight colors without creating any collision.
\end{theorem}

\section{Circle Formation in ASYNC \label{cf}}

When executing the previous algorithm, the agents    reach a configuration $\mathcal{C}_{{final}}$ in which all agents are corners of $\mathcal{H}_{{final}}$. Starting from this particular  configuration   it is possible to arrange the agents in such a way to reach a configuration $\mathcal{C}_{{circ}}$ in which each agent is positioned on the $SEC(\mathcal{C}_{{final}})$.  
Note that the solution of the {\sc Complete Visibility} problem when $|{\cal V}_0| = 2$ already form a circle, hence we focus on the case  $|{\cal V}_0| > 2$.

Notice that, when all agents are on $\partial {H}_{{final}}$ they can compute the same $SEC(\mathcal{C}_{{final}})$ since all the local views are consistent.   Moreover,  there exists a set of agents $\mathcal{X} \subseteq \mathcal{A}$ that are already on   $SEC$, and   $|\mathcal{X}| \geq 2$. 
The idea of the algorithm is to move all agents on   $SEC$ in such a way that   in each point of their trajectories they can see a subset of nodes $\mathcal{Y}$ such that $SEC(\mathcal{Y})=SEC(\mathcal{X})=SEC(\mathcal{C}_{{final}})$. More precisely, the moving rule allows agents to move towards $SEC$ if they are   ``neighbors"    (i.e., neighboring corner) of some agent on   $SEC$   in   $\mathcal{C}_{{final}}$   (see Figure \ref{rule1}). Let $a$ be neighbor of some $b$ already on $SEC$, and let $c$ be its other neighbor: $a$ will move toward $SEC$ on  line $ c  a$ 
  guaranteeing  that  the  corner-agents of the convex hull stay corner-agents, and 
do  not loose visibility with any other agent.
Note  that, unless in final position,  there is always at least one agent that can move. The algorithm terminates when all the agents are on $SEC$.

It is not difficult to see that:

\begin{theorem}\label{obsfree}
Starting from    a configuration $\mathcal{C}_{{final}}$ in which all the agents are corners,   there is an algorithm 
in ASYNC that  makes the agents    reach a configuration $\mathcal{C}_{circ}$ in which each agent   occupies a different position on $SEC(\mathcal{C}_{{final}})$ without colliding.
\label{t3}
 \end{theorem}

\newpage
    
    \paragraph{\bf Acknowledgements.} This work has been supported in part by the National Science and Engineering Research Council of Canada, under Discovery Grants, and by Professor Flocchini's University Research Chair.

\renewcommand{\headrulewidth}{0pt}

\bibliographystyle{plain}

 \end{document}